\documentclass{article}

\usepackage[utf8]{inputenx} 
\usepackage[T1]{fontenc}    

\usepackage[
 svgnames,
 table,
]{xcolor}
\usepackage[
 hyperindex,
 colorlinks,
 urlcolor = NavyBlue,
 citecolor = NavyBlue,
 linkcolor = NavyBlue,
]{hyperref}
\usepackage{authblk}

\newcommand\blfootnote[1]{
    \begingroup
    \renewcommand\thefootnote{}\footnote{#1}
    \addtocounter{footnote}{-1}
    \endgroup
}

\usepackage{lmodern}           
\usepackage[scaled]{beramono}  
\usepackage[final]{microtype}  
\usepackage{units}
\usepackage{booktabs}
\usepackage{arxiv}
\usepackage{tablefootnote}

\newcommand{\headeright}{}
\newcommand{\undertitle}{Preprint}
\newcommand{\shorttitle}{ORB: An Open Radio Buoy For Coastal Water
  Measurements}
\usepackage{tikz}
\usetikzlibrary{shapes, arrows, positioning}
\usetikzlibrary{automata, shapes.geometric, arrows, arrows.meta, calc, through}
\usepackage{pgf-umlsd}

\usepackage{url}            
\usepackage{booktabs}       
\usepackage{amsfonts}       
\usepackage{nicefrac}       
\usepackage{lipsum}		
\usepackage{graphicx}
\usepackage{natbib}
\usepackage{doi}
\usepackage{todonotes}
\usepackage{bm}

\usepackage{hyperref}

\usepackage{amsmath}
\usepackage{outlines}

\DeclareRobustCommand{\uvec}[1]{{%
		\ifcsname uvec#1\endcsname
		\csname uvec#1\endcsname
		\else
		\bm{\hat{\mathbf{#1}}}%
		\fi
}}

\usepackage{graphicx} 

\graphicspath{{figures/}}

\newcommand{\sensorReadingsPerPacket}{10}
\newcommand{\sensorDefaultStorageValue}{50}
\newcommand{\sensorDefaultMeasurementPeriod}{15}
\newcommand{\defaultBuoyReceptionFrequency}{863}
\newcommand{\defaultReadingsBeforeSending}{4}
\newcommand{\defaultTransmissionPeriod}{30}

\newcommand{\PCBLength}{63.3}
\newcommand{\PCBWidth}{35.2}
\newcommand{\PCBHeight}{34.2}

\title{ORB: An Open Radio Buoy\\ for Coastal Water Measurements}

\author[1,*]{
  Lars Willas Dreyer}
\author[2]{Andrea Pferscher}
\author[2]{Riccardo Sieve}
\author[3]{\\Jean~Rabault}
\author[1]{Atle Jensen}
\author[2]{Einar Broch Johnsen}
\author[4]{Gaute Hope}

\affil[1]{Department of Mathematics, University of Oslo, Norway,
\texttt{larswd@proton.me}, \texttt{atlej@math.uio.no}}
\affil[2]{Department of Informatics, University of Oslo, Norway,
\texttt{\{andreapf,riccasi,einarj\}@ifi.uio.no}}
\affil[3]{IT Department, Norwegian Meteorological Institute, Norway,
\texttt{jean.rblt@proton.me}}
\affil[4]{R\&D Department, Norwegian Meteorological Institute, Norway,
\texttt{gauteh@met.no}}
\date{\today}

\begin{document}

\newcommand{\revised}[1]{\textcolor{black}{#1}}
\maketitle

\begin{abstract}
Oceanographic instrumentation technology is currently under rapid transition towards increasingly open-source technology. Open-source buoys compete with commercial and closed-source buoys both in price, functionality and availability. Long-range radio (LoRa) is a communication technology which is inexpensive both in terms of data transfer cost and power without the need for \revised{pre-existing} infrastructure. In this paper, we present ORB, an open-source drifter buoy using LoRa for coastal water measurements. ORB is designed to be reliable, low-cost, modifiable and power efficient.  We present validation experiments demonstrating that ORB can achieve a radio telemetry range of \revised{five kilometers}, and has an expected battery lifetime of up to seven months\revised{, owing largely to a the radio telemetry which operates at the low power consumption of \unit[7]{mA}}. Finally, we discuss the role and contribution of ORB in the space of open-source instrumentation and ocean modeling.
\end{abstract}

\blfootnote{${}^*$ Corresponding author.}

\section{Introduction}
\label{eq:intro}

One of the most important mechanisms regulating oceanic dynamics and ocean health is ocean circulation. These dynamics are active from the global scale, such as the Gulf-stream \citep{palter2015role}, to the local scale, such as fish egg transport \citep{sundby2015principles, norcross1984oceanic, https://doi.org/10.1111/fog.12474}, pollutant transport such as oil spill drift \citep{le2012surface}, and coastal nitrate deposition \citep{sugimoto2009transport,sigleo2005nitrate}. Coastal environments are especially vulnerable to human activity \citep{roukounis2022indices}. Increased precipitation due to climate change might result in increased river erosion \citep{toimil2017managing}, and the same rivers also constitute one of the main pathways for industrial and agricultural pollution by humans \citep{vikas2015coastal, howarth2000nutrient}. High-quality measurements of oceanographic flow are essential to understand ocean dynamics and ocean health.

Measuring the ocean has historically been a challenging task, as researchers had to perform  measurements manually and often \emph{in-situ} (e.g., \citep{zielinski2021history}). Satellite imaging has drastically increased our ability to measure and model the ocean since the first satellite was launched in 1977 \citep{freeman2010ocean}. Today, we can get daily measurements of the global sea surface at the kilometre scale \citep{amani2022ocean}. However, this resolution alone is often insufficient to capture important information about flow, especially in scenarios where the important length scales are smaller than those of the open deep sea. 

An alternative to in-person measurements is sensor buoys, which allow us to measure the system of interest in-situ. Such instruments have been commercially available since the 1960s \citep{datawell2006datawell, datawell2011history}. These instruments have historically been priced in the range of 10-100 thousand USD \citep{rabault2020open}, and been closed-source hardware with limited customizability. The price of a sensor buoy fell in the early 2000s with increased competition from other commercial buoys \citep{raghukumar2019directional} and closed-source academic buoys \citep{da2022generation, kohout2015device, wilkinson2007autonomous}. The availability and reduced price  of these buoys made it much more feasible to collect larger amounts of data of the ocean and measure ocean conditions for which in-person measurements are too difficult or dangerous to perform. 

Several prognostic models of ocean flow have been developed to understand phenomena such as oil spills \citep{perianez2020lagrangian,korotenko2010high}, the dispersion of brine from desalination \citep{pereira2021brine, wood2020modelling}, and the transport of submerged bodies (such as humans at sea) \citep{hackett2006forecasting, tu2021predicting}. However, these ocean models still suffer from a lack of data, either in sufficient resolution or in its entirety \citep{fringer2019future, cavaleri2025more}. In a recent workshop in Oslo in March 2025, the need for an increased amount of sensor data for forecasting in general, and the arctic in particular, was an ongoing topic of discussion amongst representatives from many of the meteorological agencies across Europe and North America \citep{muller25conference}. 

Recently, several low-cost open-source instruments have been released. These make the process of taking oceanographic measurements easier, less expensive, and more customizable. Notable examples include the \emph{Small Friendly Buoy} (SFY) \citep{hope2025sfy} to measure waves close to the coast, the \emph{OpenMetBuoy} (OMB) \citep{rabault2022openmetbuoy} to measure wave activity under arctic and open ocean conditions, and the \emph{microSWIFT} buoy \citep{thomson2024development} which is capable of measuring wave activity, temperature, and salinity and is deployable from an aircraft. These sensor buoys are part of a recent trend: a shift from commercial, closed-source, and expensive buoys toward cheaper, open-source, and scientist-developed instrumentation. The open-source aspect in particular is an important development, because openly available source code serves the dual purpose of both allowing the users to tailor the instrument to their specific use case and simultaneously opening up the entire software implementation to external scrutiny. This transparency allows any user to check that the framework works as stated, while reducing the chance for undiscovered error in the firmware to persist indefinitely. 

Most of these buoy designs utilize a GSM or an Iridium satellite communication solution to transmit buoy measurement to the end user. Iridium uses satellite communication \citep{maine1995overview} and therefore offers a reliable connection even in distant or inhospitable environments such as the Arctic \citep{rabault2022openmetbuoy}, while GSM uses the cell phone network to transmit possibly large amounts of data quickly \citep{hope2025sfy}. However, both solutions consume large amounts of power (communication is the main power sink for both the OMB and the SFY buoys) and may come with significant cost in data transfer fees (in particular when using Iridium). One way to circumvent these limitations is to use low-power communication protocols such as LoRa \citep{devalal2018lora}. LoRa is a free band radio technology that transmits radio signals at a sub-GHz frequency with low power consumption, over a range of up to a few kilometres. Compared to GSM and Iridium, LoRa is more limited in range, but offers a cheap and power-efficient mode of data transfer in regions or applications where the distance to a receiving node can (to a certain extent) be controlled. Several remote sensor projects have used LoRa as their communication protocol. The drifter buoy of \cite{majumder2024buoys} uses LoRa together with an \emph{Internet of Things}-network, while \cite{raphael2025low} use a network of LoRa devices surrounding a central base station with a satellite communication setup. However, neither of these buoy designs is open source, making both buoys difficult to adapt according to specific needs. 

The paper presents ORB,  a new open-source LoRa-based drifter buoy and base station framework made for measuring circulation and coastal currents. This buoy framework offers a simple, reliable and inexpensive way to measure coastal flow and ocean dynamics, with a battery lifetime of several months. Both the default code and hardware configurations are publicly available on GitHub, and come with a modular setup which can readily be extended to include additional functionality. Furthermore, the ORB buoys are relatively small and light, allowing field experiments to be performed without the need for more specialized transportation than a regular car, as well as ensuring that the base station installation is largely nonintrusive.  We report on validation data from both radio range experiments and a full-system reliability experiment, demonstrating a battery lifetime of more than three months.

We expect that the open-source implementation of the ORB will \revised{complement the already rich toolbox of open-source instrumentation currently available to researchers. The ORB is in particular crafted to offer a long-life, even more budget friendly option than other pre-existing buoy solutions when the range of the experiment is within the coverage of the LoRa telemetry.} A large deployment of buoys will, for example, yield a data ensemble of the oceanographic flows in coastal environments at a fraction of the cost of commercial alternatives. These data ensembles can be used to validate numerical models, feed information to digital twins, and other actors interested in  coastal climates, all with high spatial and time resolution in the area of interest. Furthermore, by providing an \emph{adaptive measurement period}, the ORB is able to sample areas of high activity with higher temporal resolution, such as rivers, estuaries, and local currents. 

\section{ORB: An Open Radio Buoy Framework}
\label{sec:ORBoveriew}

\begin{figure}[t]
    \centering
    \includegraphics[width=0.55\linewidth]{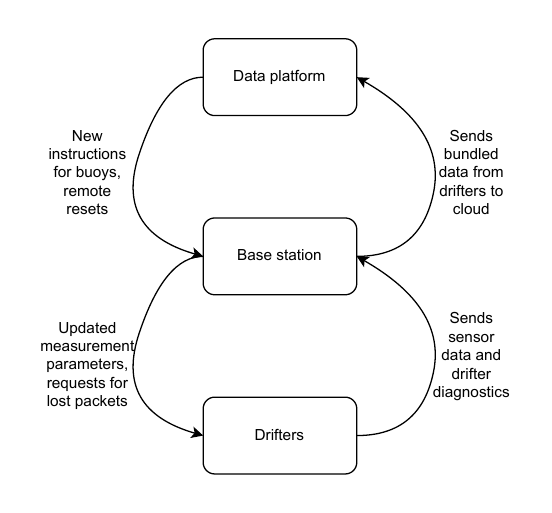}
    \caption{Relationship diagram showing how the different parts of the ORB framework interact. Drifters are deployed in a coastal area within range of one or more base stations. They gather in situ measurements, which are sent via LoRa to a base station, which then bundles the data and sends them to the backend data platform over GSM. Conversely, the data pipeline allows the user to remotely give instructions to both the base stations and the drifters by sending instructions from the backend data platform to the base stations, and from there to the buoys.\label{fig:relationship_diagram}}
\end{figure}

ORB is both a buoy and a framework for monitoring coastal water environments. The framework consists of a drifter buoy design, a base station design, and a visualization/data platform design. The relationship between the different parts of the framework is visualized in Figure \ref{fig:relationship_diagram}.

In our default configuration, the drifters are surface drifters equipped with a GPS, one or more thermometers, an SD card, and a LoRa transceiver. In addition, more sensors can be attached by the user through a variety of interfaces (I2C, SPI, serial, analogue). The LoRa transceiver also acts as the main control unit of the drifter. The default drifter firmware is customizable through a configuration file, with important parameters explained in Tables \ref{tab:config_params} and \ref{tab:config_flags} (these are detailed in the following sections). Hence, the ORB drifter is quite configurable even for a user with limited programming knowledge, and it is adaptable to the needs of the end user. In the GitHub repository, example codes show how it is possible to use each component of the drifter separately, in case a user would prefer to write their own firmware from scratch. The base firmware workflow of the buoy consists roughly of two phases after the initial setup phase. The first phase is the measuring phase, where the buoy reads each attached sensor a total of $n_m$ times every $T_m$ minutes (as shown in Table \ref{tab:config_params}), before filtering and averaging all the readings into a single mean measurement value for each sensor. In he second phase, the buoy will start looking for a base station to transmit data and to receive for new instructions. This phase occurs after both the time limit $T_t$ between measurements and the minimum measurement threshold $N_m$ have been exceeded. A detailed overview of the included firmware is given in Section \ref{sec:firmware}. A diagram of the protocol is shown in Figure \ref{fig:protocol_diagram}.

The second component in the ORB framework is the base station, which acts as an intermediary between the user and the buoys, as it transmits data from the buoys to a cloud backend and is additionally equipped with the functionality necessary to remotely push setting updates to the buoy firmware during deployment. The base station is controlled by the same main controller unit and radio transceiver as the drifter, but unlike the drifter, it is not equipped with any sensors. In addition, the base station is equipped with a GSM modem, to allow communication with cloud-based servers. The base station is designed to have a long battery life and to be mounted at a coastal vantage point with GSM coverage, close to the area where the buoys are deployed. Remark that the LoRa radio protocol has reduced range if there is no line of sight between the sender and receiver, so the user must take some care when placing base stations in the field.

\begin{table}[t]
    \centering
    \begin{tabular}{@{}p{43mm}@{\hspace{-2mm}}ccp{73mm}@{}}\toprule
         \textbf{Buoy parameter} & \textbf{Default value} & \textbf{Remote update} & \textbf{Notes}\\
         \midrule Reception frequency $f_r$  & 
         \unit[\defaultBuoyReceptionFrequency]{MHz} & No & Legal bands subject to local jurisdiction \\
         Beacon frequency $f_b$ & \unit[868]{MHz}  & No & Legal bands subject to local jurisdiction \\
         Measurement period $T_m$ & \unit[\sensorDefaultMeasurementPeriod]{min} & Yes & - \\
         Transmission period $T_t$ &  \unit[\defaultTransmissionPeriod]{min} & Yes & Sets an upper bound on the transmission period \newline if measurements per transmission is low \\
         Beacon message period $T_b$ & \unit[3]{min} & No & -  \\
         Measurements per \newline transmission $N_m$ & \defaultReadingsBeforeSending & Yes & Sets an upper bound on the transmission period\newline if transmission period is low. \\
         Max number of measurements & \sensorDefaultStorageValue & No & - \\ stored in memory, $N_\text{max}$ &  &  &  \\
         Readings per measurement $n_m$ & 15 & No & Higher number of readings for a single \newline  measurement yields better filtered estimate \\
         Max GPS read time $T_\text{GPS}$ & \unit[3]{min} & No & Max time the GPS will spend looking for a fix. \\
         LoRa bandwidth $B_L$ & \unit[125]{kHz} & No & -\\
         LoRa spread factor $s_f$ & 8 & No & Must be an integer between 5 and 12 \\
         LoRa coding rate $c_r$ & 6 & No & Must be an integer between 5 and 8 \\
         LoRa output power $W_\text{LoRa}$ & \unit[20]{dBm} & No & Must be an integer between -9 and 22 \\  
         Target distance between\newline  measurements $d_\text{target}$ & \unit[200]{m} & Yes & For the adaptive measurement period  \newline described in Equation \ref{eq:f_adaptive} \\
         Velocity threshold $v_t$ & \unitfrac[1.0]{m\,}{\,s} & Yes & Threshold for the adaptive measurement\newline algorithm to activate. \\
         \bottomrule
    \end{tabular}
    \smallskip
    \caption{The most important parameters for the ORB buoys, all of which are configurable by the user in the configuration file, their default values, whether they can be changed remotely and important notes about the parameters.    \label{tab:config_params}}
    \smallskip
    \begin{tabular}{@{}p{35mm}lp{100mm}@{}}
         \toprule \textbf{Buoy functionality} & \textbf{Default setting} & \textbf{Notes}\\
         \midrule Remove outliers  & Enabled & Enable/Disable $\sigma$-filtering \\
         Enable motion detection & Enabled &  Adaptive measurement period
         as detailed in Equation \ref{eq:f_adaptive}\\
          Perform handshake &  Enabled & Toggle handshake portion of Figure \ref{fig:protocol_diagram}. \\
         Log every reading  & Enabled & Toggle if the ORB should store all readings to SD card,\newline or just filtered mean. \\
         Transmit deployment\newline message & Disabled & Enable to ensure that the buoy will send an initial message\newline to a nearby base station in range\newline
before entering the main program loop. \\
         \bottomrule
    \end{tabular}
    \smallskip
    \caption{The most important flags which can be enabled or disabled by the user to activate or deactivate parts of the ORB functionality.\label{tab:config_flags}}
\end{table}

The third component of the ORB framework is a visualization/data platform. The data platform offers a web page frontend that allows monitoring the deployed buoys, visualizing and inspecting  the collected data, and issuing new instructions to the buoys via the deployed base stations. The data platform is, unlike the base stations and buoys, not strictly necessary for the framework to operate, as the GSM module in the base station transmits directly to the online data repository; we here use the \url{notehub.io} data platform maintained by Blues, the manufacturer of the base station GSM module. The ORB data platform is a front-end made to provide easier and more intuitive interaction with the buoys during deployment.

\section{The Design of the ORB Buoy}
The ORB buoy is the data collection component of the ORB framework. It consists of a main controller unit with a radio transceiver, a GPS, an SD card reader/writer and any additional measuring devices installed by the end user. In this section, we detail the software, hardware and functionality of the ORB buoy. 

\subsection{The Main Controller}
The main controller unit's primary job is to gather readings from the attached sensors, post-process these readings and then store them either in RAM or in local storage until the data can be recovered. Furthermore, the main controller should also facilitate the data transfer protocol between the drifter and base station. 

The main controller unit of the ORB buoys is the Wio-E5 Mini board by \cite{wioE5mini}. The Wio-E5 Mini board was chosen as the main board for the ORB buoys in part due to the price ($\sim$ 20 USD) and in part because it incorporates a low-cost high-performance main controller unit that has a built-in LoRa transceiver \citep{wioE5mini}. This transceiver is the sole communication unit of an ORB buoy, and transmits any data to a nearby base station if there are any. In the case where no base station is present within LoRa range, the buoy will try again at a later time (default 15 min). Each Wio-E5 Mini board has a 12 byte unique identifier number, which we use as the identifier of the associated buoy.  The LoRa protocol, being very low power, has a limited range, and we have observed ranges of up to 2 km in ocean waters, but ranges of up to 10 km have been observed in more ideal conditions \citep{aref2014free, falanji2022range}. During the setup of a field experiment, we were able to attain a range of up to 5.4 km while the buoy was on land.

The Wio-E5 Mini board uses a low-cost high-performance STM32WL55JC microcontroller unit, with a 32-bit \unit[42]{MHz} processor, 64 kb RAM and comes with a built-in LoRa module. Furthermore, the STM32WL55JC comes equipped with a wide range of interfaces (several I2C, SPI, serial buses, as well as digital and analogue pins), facilitating an easy way of adding additional sensors and functionality to the buoy, while still leaving a significant amount of embedded memory and compute left for the end user. The firmware uploaded to the Wio-E5 Mini board is programmed using Platformio IDE, and is uploaded to the microcontroller via a ST-LINK debugger, and the firmware is openly available online.\footnote{\url{https://github.com/larswd/OpenLoraBuoy}}

\subsection{Sensor Components}
The ORB buoys are additionally equipped with a thermometer and a GPS. The GPS measurements ensure both a spatial and temporal location for the other sensor readings. In the default software, a faulty GPS will result in a firmware freeze, to minimize the risk of deploying a faulty drifter. The thermometer is a cheap and power efficient instrument which allows the buoy to capture surface temperatures, or temperatures further down the water column. These values are important parameters in many operational models in meteorology, such as GOTM \citep{umlauf2005second} or ROMS \citep{moore2011regional}. An overview of all components of the ORB buoy, their function, and a link to the page where they were ordered, together with prices, is given in Table \ref{tab:components}, and the wiring of these is shown in Figure \ref{fig:schematic}.

Each buoy is, in the default configuration, equipped with a single DS18B20 \citep{DS18B20} thermometer, which uses a one-wire protocol to transmit measured temperature data to the main controller unit. The thermometer has an operational range of -10--85 degrees centigrade, with an absolute accuracy out of the box of $\pm0.5$ degrees centigrade. However, the dynamic accuracy is significantly higher and the repeatability and stability of the sensors allow custom calibration and higher accuracy. Specifically, the thermometers can be calibrated with a three-point calibration, improving the absolute accuracy to approximately 0.1 degrees centigrade (for more information, see \citet{muller2024svalbard,muller2025svalbard2024}). Furthermore, the DS18B20 can be daisy-chained to enable measurement of a temperature profile. DS18B20 sensors are calibrated prior to deployment and the calibration is applied as a post-processing step, as the buoy does not store calibration data in memory, meaning an a-posteriori lookup table which matches thermometer to calibration curve is necessary to get the most accurate temperature readings. While post-deployment calibration is possible, we advise against it due to the risk of not recovering the buoy after deployment. 

\begin{table}[t]
    \centering
   \begin{tabular}{@{}p{42mm}lp{22mm}@{}r@{}}
         \toprule \textbf{Component} & \textbf{Function}  & \textbf{Producer} & \textbf{Price (per 08-2025) [USD]}\\
        \midrule Wio-E5 Mini & Main controller unit, radio transceiver & \href{https://wiki.seeedstudio.com/LoRa_E5_mini/}{Seeed studio} & 21.90 \\
        Adafruit Ultimate GPS  & GPS sensor, RTC & \href{https://www.adafruit.com/product/746}{Adafruit} & 29.95$^*$\\
        DS18B20 & Thermometer & \href{https://www.sparkfun.com/temperature-sensor-waterproof-ds18b20.html}{Sparkfun} & 10.95 \\
        Micro-SD card\newline breakout board+ & SD card writer/reader & \href{https://www.adafruit.com/product/254}{Adafruit} & 7.50 \\
        Pololu 3.3V  S7V8F3 & Step up/down voltage regulator & \href{https://www.pololu.com/product/2122}{Pololu} &9.95$^*$ \\
        SD card & On buoy data storage & \href{https://www.adafruit.com/product/1294}{Adafruit} & 9.95 \\
        Saft LSH20 & 13 Ah Li-battery & \href{https://www.nkon.nl/en/saft-lsh-20-lithium-battery-3-6v.html}{Saft} & 18$^*$ \\
        PCB    & Base circuit board &\href{https://jlcpcb.com/}{JLCPCB} & 1-6$^*$ \\
        Container parts & Waterproof buoy container equipment  & - & $\sim$15-20  \\ 
        \midrule
        \multicolumn{2}{@{}l}{Price per buoy for 5 buoys}  & & 126.2 \\
        \multicolumn{2}{@{}l}{Price per buoy for 50 buoys} &  & 115.9 \\
        \bottomrule
   \end{tabular}
   \smallskip
    \caption{The hardware components of the ORB buoy, their function and manufacturer. For convenience, manufacturer names are  hyperlinked to the product page where the components were ordered. Prices marked with an asterisk ($^*$) are subject, at the time of writing, to a bulk order discount. In particular, the cost of a PCB is almost exclusively shipping, and the price listed is a price estimate for a single PCB out of an order of 5--30 PCBs. Hence, the final price might be lower than what is listed here for several of the components.\label{tab:components}}
\smallskip
    \begin{tabular}{@{}p{67mm}c@{\qquad}c@{\qquad}c@{}}
        \toprule \textbf{Phase} & \textbf{Current} & \textbf{Phase Duration} & \textbf{Power consumption per day} \\
         \midrule Inactive mode & \unit[0.7]{mA} & $\sim$\unit[15]{minutes} & \unitfrac[17]{mAh\,}{\,day}\\
         Measuring mode (GPS) & \unit[45-60]{mA} & \unit[10-60]{seconds}\footnote{During experiments, typical time in this phase is 12 seconds, although 60 seconds has been seen.} & \unitfrac[40]{mAh\,}{\,day} \\
        Measuring mode (Other) & \unit[8-10]{mA} & \unit[10]{seconds} & \unitfrac[2]{mAh\,}{\,day} \\
         Transmission mode & \unit[9-11]{mA} & \unit[1]{minute} & \unitfrac[12.8]{mAh\,}{\,day} \\
         \hline Total consumption: bad conditions & - & - &  \unitfrac[123]{mAh\,}{\,day}\\
         Total consumption: moderate conditions & - & -  & \unitfrac[67]{mAh\,}{\,day}\\
         Total consumption: good conditions & - & - & \unitfrac[47]{mAh\,}{\,day} \\
         \bottomrule
    \end{tabular}
    \smallskip
    \caption{The power consumption of the ORB buoy during each distinct phase of the firmware execution, the typical duration of each phase, and the amount of power consumed by each phase during a day with moderate conditions. Here, \emph{good conditions} happen when the GPS is quick to get a fix and pulls a low amount of power (average of 50 mA for 10 seconds), \emph{moderate conditions} when the power consumption of the GPS and time is slightly larger (60 mA for 25 seconds), while \emph{bad conditions} occur at the high end of the GPS power use (60 mA for 60 seconds). For the other modes, we assumed a current of 9 mA in the measuring mode (other) phase, and a current of 10 mA in the transmission phase.\label{tab:power_usage}}
\end{table}

The GPS measurements have a precision of 3 meters, which is a hard limit of the instrument \citep{AdafruitGPS}. During a measurement cycle, the GPS is always read first as the GPS fix information is necessary both for spatial localization of measurements and to update the real time clock of the drifter.  A single GPS data reading contains a value for latitude, longitude, speed and direction, as well as a (POSIX) timestamp and a reading ID. This ID allows for remote recovery of lost data through instructions transmitted to the buoy from the base station.  

The open-source firmware and hardware in the ORB buoy allows for user modifications, both in terms of minor software adjustments, such as changes to the measurement and transmission protocol as well as  major additions such as new hardware. The default firmware contains examples of how to add additional sensor firmware. The PCB has a second row of pins for each pin on the ORB. Furthermore, the firmware is designed in a containerized fashion, meaning the radio transceiver and SD card logger is applicable to new hardware. In the GitHub repository, we present an example of how to add a turbidity sensor to the ORB. 

\begin{table}[t]
    \begin{tabular}{@{}p{25mm}c@{\qquad}c@{\qquad}c@{\qquad}c}
        \toprule \textbf{Buoy technology} & \textbf{Instrumentation} & \textbf{Battery lifetime} & \textbf{Cost (per buoy/month)} & \textbf{Telemetry}\\
         \midrule ORB & GPS, Temperature & \unit[3-7]{months} & \unitfrac[120]{USD\,}{\,buoy} & LoRa \\ & & & - & \\ \\
         SFY  & GPS, IMU & \unit[12]{days} & \unitfrac[200]{USD\,}{\,buoy} & GSM \\ \citep{hope2025sfy} & & & - & \\ \\
        OMB  & GPS, IMU & \unit[4.6]{months} & \unitfrac[560]{USD\,}{\,buoy} & Iridium \\ \citep{rabault2022openmetbuoy}& & & + \unitfrac[110]{USD\,}{month\,} & \\ \\
         microSWIFT & GPS, IMU, Temperature & \unit[2-5]{days} & \unitfrac[900]{USD\,}{\,buoy} & Iridium \\ \citep{thomson2024development}& & & Not given & \\ \\
         LainePoiss & GPS, IMU$^{[\text{a}]}$ & \unit[2]{months} & \unitfrac[5500]{USD\,}{buoy\,} & GSM, Iridium\\ \citep{alari2022lainepoiss} & & & Not given & \\ \\
         \citet{yurovsky2020mems} & IMU & \unit[15]{hours} & Not given & None$^{[\text{b}]}$\\ & & &  & \\
         Spartacus & Camera, GPS & Weeks$^{[\text{c}]}$ & \unitfrac[5000]{USD\,}{buoy\,} & GSM \\ \citep{cavaleri2025more}  & & & - & \\ \\
         (X)FZ & GPS, IMU & \unit[3]{weeks} & Not given & Iridium\\ \citep{kodaira2024affordable} & & & Not given & \\ \\
         MELODI & GPS,IMU$^{[\text{a}]}$, Temperature & unlimited$^{[\text{d}]}$ & \unitfrac[2000]{USD\,}{buoy\,} & GSM, Iridium \\ \citep{mironov2024miniaturized} & wind$^{[\text{f}]}$ & & +\unitfrac[140]{USD\,}{month\,} & \\ \\  
         OWL\citep{rabault2026openwavelogger}& GPS, IMU & \unit[20]{days} & \unitfrac[220]{USD\,}{buoy\,} & None$^{[\text{e}]}$ \\
         \bottomrule
    \end{tabular}
    \caption{A non-exhaustive list of open source instruments for oceanographic measurements comparing their instruments, battery lifetime, cost and telemetry options. Here, GPS is short for Global Positioning system, IMU is short for Inertial Measurement Unit, GSM is short for Global System for Mobile Communications and Iridium is a satellite communication protocol. The price and battery lifetime listed is based on the reported values at publication, and can have been changes since. Footnotes: [a]: 2D wavespectrum, [b]: deployed from a fishing rod, no telemetry needed, [c]: Battery lifetime is given as weeks to months, but no specific number is given, [d]: Solar panel powered, [e]: only logs to SD card, [f]: in development.   \label{tab:buoy_alternatives}}
\end{table}
\subsection{Additional Components}
In addition to the sensors, the ORB buoy comes equipped with several supporting components: a battery, a voltage regulator and an SD card writer. The battery used in our tests is a single SAFT LSH20 3.6\,V lithium D battery, with a capacity of \unit[13]{Ah}. The output power is regulated by a 3.3\,V step-up/step-down regulator \citep{pololu3.3V}. We measured the power usage of the buoy with a multimeter connected in serial between the voltage regulator and the reed switch. Using the currents measured by the sensor (see Table \ref{tab:power_usage}), and assuming a full \unit[13]{Ah} battery,  the buoy has an expected lifetime between  3.5 months (worst case) to 9.2 months (best case) of continuous operations, although more rapid measurements and transmissions would decrease the expected lifetime.

The ORB buoy transmits the recorded data and saves it to an SD card via an Adafruit microSD breakout \citep{AdafruitSD}. The ORB buoy sleeps for a certain period of time between measurements. The default number of measurements during a given measurement cycle is \sensorReadingsPerPacket, but this number is configurable in the code. The buoy transmits the filtered (as explained in Section \ref{sec:firmware}) average of these \sensorReadingsPerPacket \, measurements as an estimated value for the corresponding measurement cycle. Thereafter, a new mean of all the remaining readings is computed and stored in memory for transmission.  However, each individual measurement is saved to the SD card. Hence, if the ORB buoy is recovered, the raw data could be analysed in more detail.


\begin{figure}[t]
    \centering
    \input{figures/figure2/PCB_schemantic}
    \caption{Subfigure (A) shows a schematic for the wiring between each component on the ORB buoy. The reed switch enables the user to turn off the buoy by placing a magnet on a designated spot on the buoy casing. The pull-down resistance between the EN-pin on the GPS and ground is to ensure that the GPS is switched off while not measuring. A small coin cell battery is connected to the GPS to allow it to store fix information even when powered off. Subfigure (B) shows the corresponding printed circuit board (PCB), where the connections depicted in Subfigure (A) are implemented.\label{fig:schematic}}
\end{figure}

\subsection{Default Software}
\label{sec:firmware}
ORB is an open-source buoy and hence programmable by the end user to tailor it to a specific use case. It comes with a default software to read, record, and transmit data from all sensors connected to the main controller unit. The default software can be roughly divided into two main stages (\emph{setup} and \emph{measurement and transmission} modes), the latter can be further divided into four  stages. The software includes a configuration file in which many parameters relating to buoy functionality, such as measurement frequency, transmission frequency, SD-logging, and motion detection, can be set by the user without the need to alter the software or code any extra functionality. In particular, we note that the frequency bands that are open for public use vary by region, and we have listed the largest different bands in Table \ref{tab:LoRabands}. The list of the most important parameters and options present in the configuration file is shown in Tables \ref{tab:config_params} and \ref{tab:config_flags}. Each named variable in this section is listed in these two tables, with the exception of the variables in Equation \ref{eq:f_adaptive}.

\begin{table}[t]
\centering
   \begin{tabular}{@{}p{30mm}ll}
         \toprule \textbf{Frequency plan} & \textbf{Frequency band} & \textbf{Areas}\\
        \midrule EU868 &  \unit[863-870]{MHz} & EU, EEC $^*$, most of Southern Africa, the Philippines, \\ & & Saudi Arabia, The United Kingdom, The Vatican \\
        CN779    & \unit[779-787]{MHz} & The People's Republic of China \\
        US915    & \unit[902-928]{MHz} & Most of the Americas \\
        AU920    & \unit[915-928]{MHz} & Australia, Argentina,  Brazil, Chile, Ecuador, New Zealand \\
        AS920    & \unit[920-923]{MHz} & Singapore, Japan, Malaysia, \\
        AS923    & \unit[923-925]{MHz} & Brunei, Cambodia, Hong Kong, Indonesia, Laos, Taiwan, Thailand, Vietnam \\
        IN865    & \unit[865-867]{MHz} & India \\        
        \bottomrule
    \end{tabular}
    \smallskip
    \caption{The publically available LoRa frequency bands compatible with the ORB, as accodring to the Things Network. See \url{https://www.thethingsnetwork.org/docs/lorawan/frequencies-by-country/} for an exhaustive list.   \label{tab:LoRabands}}
\end{table}

The ORB buoy begins in the \emph{setup} mode, in which it first initializes its connections with each connected hardware component, and where it boots up the radio for the first time. Afterwards, the buoy checks that both the radio and GPS are working as intended. If one of these components fails, the buoy will not start logging data. If the transmit deployment message flag is set to true, the buoy will not continue into the main software loop before a connection has been established to a base station. The buoy has a limited capacity for storing excess measurements, which can be tweaked with the $N_\text{max}$ option in the configuration file. This limit depends on the amount of functionality added to the buoy, but as all memory is statically allocated,  the compiler will not compile the program if the memory limit is exceeded. In case the buoy is unable to find a base station until the controller memory is full, the oldest measurement is dropped from memory. The dropped readings can be recovered from the SD card on request from the base station, as each message contains an increasing measurement number to help keep track of untransmitted measurements.

Afterwards, the ORB buoy enters the \emph{measuring and transmission} mode, which is a loop that repeats until the buoy either breaks, runs out of power, or is deactivated manually after recovery. The buoy will take measurements at a fixed or adaptive rate. The default measurement frequency $f_\text{d}$, as well as the minimal measurement frequency $f_\text{min}$ and the maximal measurement frequency $f_\text{max}$ can be set by the user in the configuration file prior to deployment. During each measurement cycle, the buoy performs $n_m$ readings of each component, before averaging these values down to a single measurement value. If the \emph{remove outliers} flag is set to true, then a $\sigma$-filtering is performed, where all readings more than $\sigma_t$ standard deviations away from the mean are discarded. Furthermore, the \emph{measurement frequency} can be modified remotely via the base station. We also provide a rudimentary adaptive measurement frequency for the buoy: If this feature is activated, the measurement frequency $f$ is computed as a function of the GPS-measured speed $v$ according to the following formula:
\begin{equation}
    \frac{1}{f} = \begin{cases} \min\left(\frac{d_\text{target}}{v}, \frac{1}{f_\text{min}}\right), & v > v_t, \\
                                \frac{1}{f_d}, & \text{else},
                    \end{cases}
    \label{eq:f_adaptive}
\end{equation}
where $v_t$ is the adaptation threshold speed and $d_\text{target}$ is the desired distance between each measurement. To avoid a buoy deadlock under weather conditions that make it difficult (or impossible) to get a GPS fix, the measurement cycle terminates if no GPS fix has been obtained within a time span of $T_\text{GPS}$. 

The buoy will attempt to send the measured data to a base station during the transmission phase. The transmission phase occurs when both the minimum transmission period $T_t$ has been exceeded and a minimum of $N_m$ measurements have been stored in memory. In this phase, the buoy will look for signals from a base station (in listening mode at the frequency $f_r$), and if found, perform a handshake to verify that it can start transmitting. Afterwards, it will transmit (at transmission frequency $f_t$) as much of its stored data as possible within the timespan allotted to it by the base station. Finally, the buoy will switch back into listening mode to receive updates from the base station.  A detailed overview of the LoRa protocol is given in Section \ref{sec:LoRa}. The configuration file includes several options to tailor the LoRa communication, including options to change the bandwidth $B_L$, spread factor $s_f$, coding rate $c_r$, output power $W_\text{LoRa}$ and transmission and receiving frequencies ($f_t$ and $f_r$ respectively). 

The buoy will also regularly transmit a beacon message. These messages include the buoy ID, and the most recently measured position and its corresponding timestamp. These messages are broadcast without a handshake and at a given frequency $f_b$, and can be received by a base station which is in recovery mode. 

At the end of the loop, the buoy powers down any components it used during the loop, going into low-power sleep mode until it is time to perform a new measurement or transmission. 

\subsection{Buoy Housing and Buoy Design}

\begin{figure}[t]
    \centering
    \input{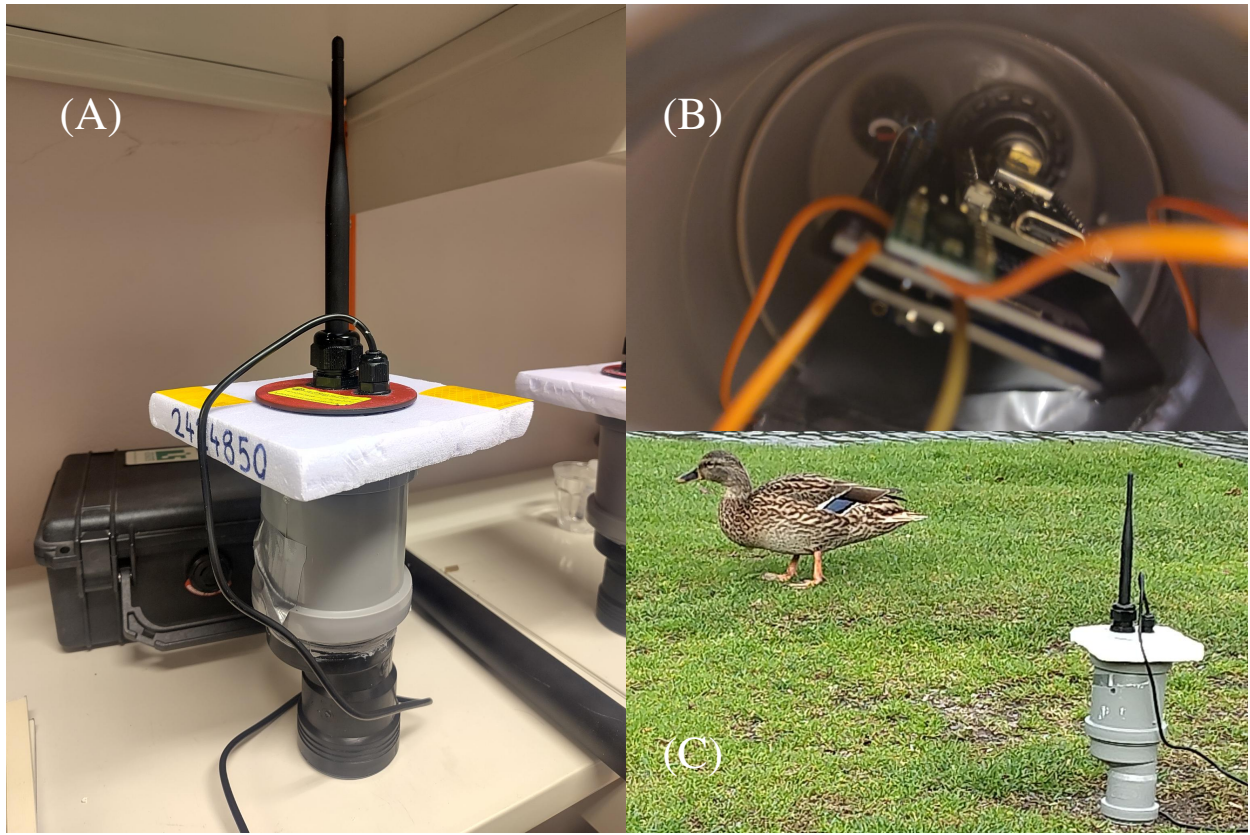}
    \caption{An assembled ORB buoy in a container with a PCB (subfigure A), and the inside of the buoy (subfigure B). Subfigure (C) shows the buoy alongside a female \textit{Anas platyrhynchos} (mallard duck) near Sognsvann, Oslo. }
    \label{fig:buoy_in_container}
\end{figure}

There is no default buoy housing for the ORB buoy. In our deployments, a buoy hull was constructed using a combination of polypropylene (PP) plastic pipes. The LoRa antenna and thermometer are led through the buoy hull using cable glands. To decide on a hull size, the minimum requirement is that it contains the printed circuit board (PCB), with dimensions \PCBLength $\times$ \PCBWidth $\times$ \PCBHeight mm. In our development, we chose to use a combination of PP plastic pipes bought at a common hardware store. The container consisted of four pipe parts, a 75 mm diameter lid, a 75 mm pipe connector, a 75 to 50 mm narrower, and a 50 mm diameter lid. In the top 75 mm lid, two holes were drilled, one hole for a cable gland for the antenna and another, smaller hole for the thermistor string. All rubber bands in the cable glands were lubed with maritime grease to reduce the risk of leakage. The narrow bottom part of the container was used as a ballast compartment. Around the top, we also fastened small styrofoam disks to ensure that the buoys were stable and always on the surface. An image of an assembled buoy is shown in Figure \ref{fig:buoy_in_container}.


An important note is the height of the antenna. The antenna included with the Wio-E5 Mini board, which we have used in the ORB buoy, is 19 cm long. As the LoRa range is severely reduced if the Fresnel zone is blocked \citep{yim2018experimental, colemanWestcottRadio}, it is important to maximize the height of the antenna. From our practical experience with the hardware assembly, the included antenna has a combining joint 4 cm above the bottom, which is a structural weak point. Having this joint above the cable gland will weaken the joint, as the cable gland will need to be as tight as possible. This weakening might cause the hull to leak and potentially a loss of the buoy. Furthermore, this is also a constraint on how low one can place a base station without problems, as placing the base station too low might lead to significant portions of the Fresnel zone being blocked by the presence of ocean waves.

\section{Base Station Components}

The second component of the ORB framework is the base station, which acts as an intermediator between the ORB drifters and the cloud, receiving messages from the drifters and uploading these to the cloud from where the data can be downloaded. The base station consists of a radio transceiver which communicates with the buoys, an SD card writer and a GSM module which transmits the received data to a cloud frontend. 

\begin{figure}[t]
    \centering
    \input{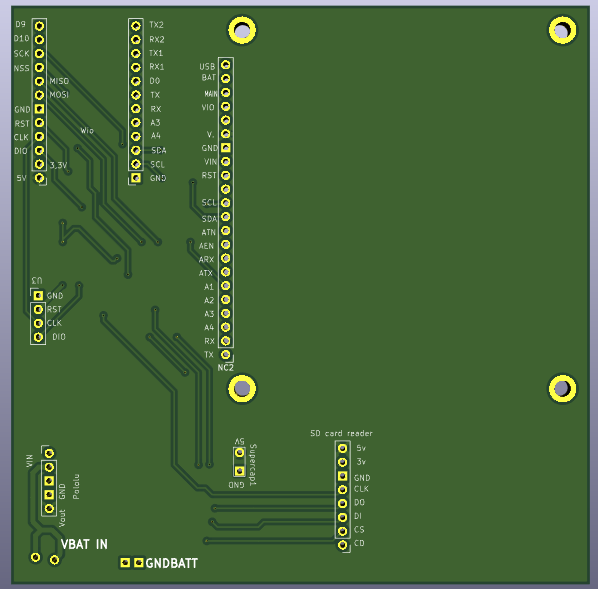}
    \caption{The wire schematic for the base station [Subfigure (A)]
      and the corresponding PCB layout [Subfigure (B)]. The Wio-E5
      Mini board acts as the radio transceiver and main control
      unit. The system operates on 5V logic, and can be reset remotely
      from the notecarrier.\label{fig:BSTSchematic}}
\end{figure}

The ORB base station is, as the buoy, controlled by a Wio-E5 Mini board \citep{wioE5mini}, which acts as main processor and LoRa transceiver to operate the base station and to receive measurements and send instructions to the ORB drifter. The base station is also equipped with a Blues Notecarrier A module \citep{bluesNoteCarrier}, which periodically transmits the received data from the buoy to the cloud  via GSM. As the data platform for our cloud frontend, we use \url{notehub.io}, which can receive, at the time of writing, 5000 messages (with a maximum message size of 1 MB) for free each month and charges 11.5 USD per 15000 messages thereafter. To minimize the amount of messages sent to the cloud, the base station stores and bundles several buoy messages into a single data packet before transmitting. Each base station is equipped with a unique base station ID, which it broadcasts at regular intervals. A schematic for the base station layout is shown in Figure \ref{fig:BSTSchematic}. 


The base station uses Notecard to enable the remote management of deployed buoys. The Notecarrier integration with Notehub allows a user to remotely send instructions to the base station, and these can be forwarded to individual buoys, as described above. For instance, a signal can be sent  to the base station to reset itself in the event of a software error. Furthermore, it is possible to request a specific measurement from a specified buoy; this request will be forwarded from the base station to the given buoy.

The ORB base station also comes equipped with local storage capabilities in the form of an SD card writer of the same kind as the one used by the buoy. As notehub, the data platform to which the Notecard sends the data, only keeps the data for a period of three days, so this local storage is a crucial fail-safe in case the data export pipeline from notehub were to fail either due to human or code error. The Adafruit SD writer works well for this purpose as it is able to operate at both a 3.3\,V logic level, used by the buoys, and at a 5\,V level,  used by the base station. In our tests, we used three 13 Ah  3.6\,V Lithium Batteries as the power source for the base station, which were regulated up to a 5\,V logic level using a Pololu step-up/step-down voltage regulator \citep{pololu5V}. A full component list with an approximate price tag can be seen in Table \ref{tab:BSTcomponents}.

\begin{table}[t]
    \centering
   \begin{tabular}{@{}p{31mm}lp{20mm}@{}r}
         \toprule \textbf{Component} & \textbf{Function}  & \textbf{Producer} & \textbf{Price (per 08-2025) [USD]}\\
        \midrule Wio-E5 Mini & Main controller unit, radio transceiver& \href{https://wiki.seeedstudio.com/LoRa_E5_mini/}{Seeed studio} & 21.90 \\
         Notecarrier & GSM module carrier & \href{https://shop.blues.com/collections/notecarrier/products/carr-al?variant=43012453171441}{Blues} & 25 \\
        Notecard  & GSM module & \href{https://shop.blues.com/products/notecard-cellular?variant=44541263118577}{Blues} & 63 \\
        Micro-SD card\newline breakout board+ & SD card writer/reader & \href{https://www.adafruit.com/product/254}{Adafruit} & 7.50 \\
        Pololu 5V  S7V8F3 & Step up/down voltage regulator & \href{https://www.pololu.com/product/4082}{Pololu} &17.95$^*$ \\
        SD card & Data storage & \href{https://www.adafruit.com/product/1294}{Adafruit} & 9.95 \\
        3 Saft LSH 20 & 13 Ah Li-battery & \href{https://www.nkon.nl/en/saft-lsh-20-lithium-battery-3-6v.html}{Saft} & 54$^*$ \\
        PCB    & Correct wire connection  between components &\href{https://jlcpcb.com/}{JLCPCB} & 1-6$^*$ \\
        Container parts & Waterproof BST container equipment  & - & $\sim$15  \\ 
        \midrule
        Price per base station &  & & 213 \\
        \bottomrule
    \end{tabular}
    \smallskip
    \caption{The hardware components of the ORB base station, their function and manufacturer. For convenience, manufacturer names are  hyperlinked to the product page where the components were ordered. Prices marked with an asterisk ($^*$) are subject, at the time of writing, to a bulk order discount. In particular, the cost of a PCB is almost exclusively shipping, and the price listed is a price estimate for a single PCB out of an order of 5--30 PCBs. Hence, the final price might be lower than what is listed here for several of the components.\label{tab:BSTcomponents}}
\end{table}

\section{Data Transmission Protocol}
\label{sec:LoRa}

\begin{figure}[h]
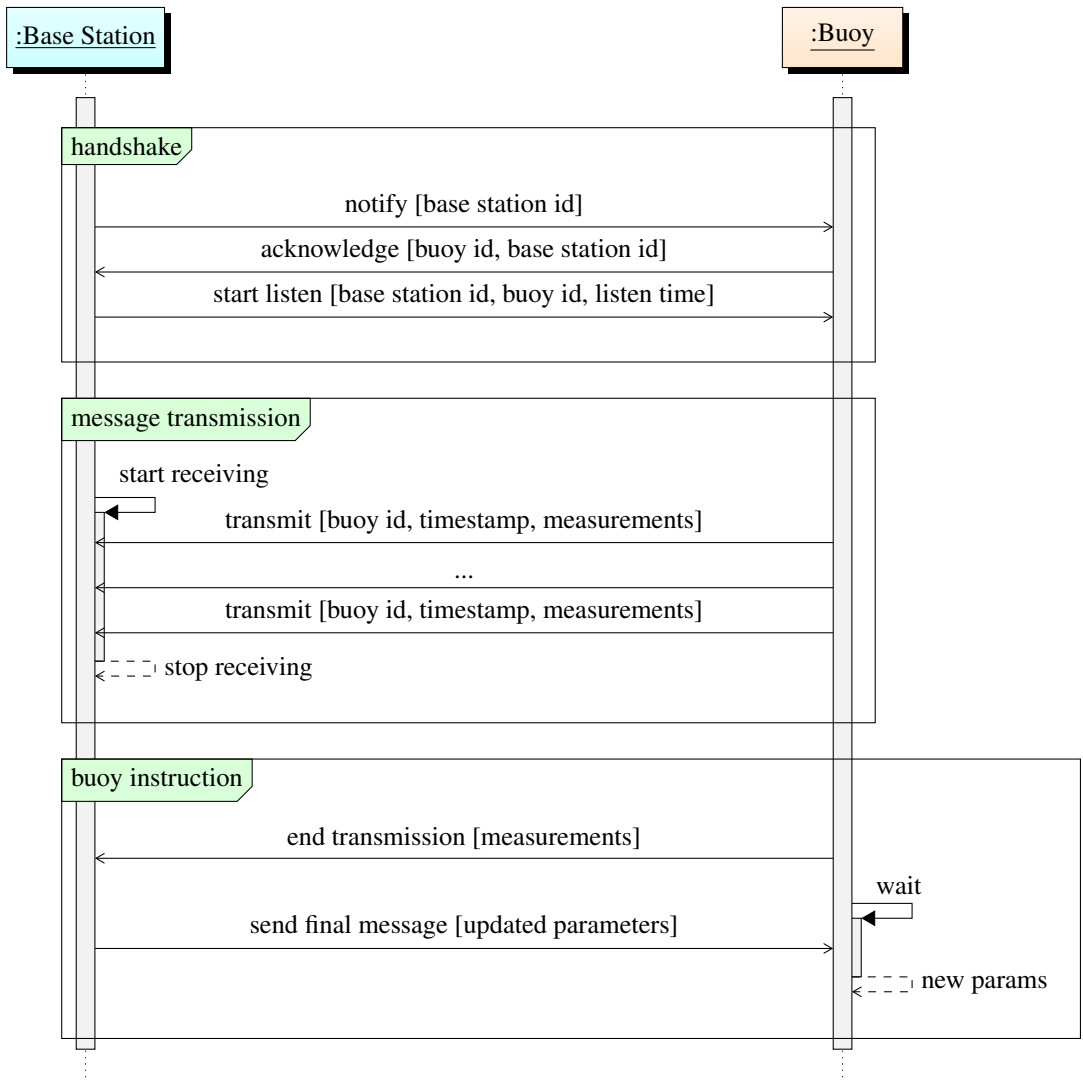

    \centering
\begin{sequencediagram}
\tikzstyle{inststyle}+=[top color=cyan!10, bottom color=cyan!20]
\newthread[gray!10]{bs}{:Base Station}
\tikzstyle{inststyle}+=[top color=orange!10, bottom color=orange!20, xshift=8cm]
\newthread[gray!10]{buoy}{:Buoy}

\begin{sdblock}[green!15]{handshake}{ }
\mess{bs}{notify [base station id]}{buoy}
\mess{buoy}{acknowledge [buoy id, base station id]}{bs}
\mess{bs}{start listen [base station id, buoy id, listen time]}{buoy}
\end{sdblock}

\begin{sdblock}[green!15]{message transmission}{ }
\begin{callself}{bs}{start receiving}{stop receiving}
\mess{buoy}{transmit [buoy id, timestamp, measurements]}{bs}
\mess{buoy}{...}{bs}
\mess{buoy}{transmit [buoy id, timestamp, measurements]}{bs}
\end{callself}
\end{sdblock}

\begin{sdblock}[green!15]{buoy instruction}{ }

\mess{buoy}{end transmission [measurements]}{bs}
\begin{callself}{buoy}{wait}{new params}
\mess{bs}{send final message [updated parameters]}{buoy}
\end{callself}
\end{sdblock}
\end{sequencediagram}
    \caption{Radio protocol between base station and buoy. The protocol is separated into three stages. The base station is, when not busy, regularly sending out a short message notifying buoys in the area that it is available for data reception. Once a buoy is ready to transmit, it will turn on the radio to listen mode and check if it gets a signal from a base station. If so, the buoy and the base station will perform a handshake to ensure that the connection is stable and to avoid multiple buoys transmitting simultaneously. Once verified, the base station enters listening mode while the buoy transmits as much data as possible within the time allotted to it by the base station. Finally, the buoy sends a message confirming it is done to the base station before switching to listening mode. If received, the base station might send updated directives to the buoy.  }
    \label{fig:protocol_diagram}
\end{figure}

The ORB buoy uses LoRa to send messages to the base station. LoRa uses radio messages on a specific frequency band (433 MHz and 868 MHz in the EU\revised{, a list of the most common LoRa bands is given in Table \ref{tab:LoRabands}), and can transmit messages for up to 10 kilometers in ideal circumstances \citep{devalal2018lora, falanji2022range}. This range is an unavoidable limit of using LoRa as the telemetry of choice, rather than GSM or Iridium, and as a result the ORB should primarily be considered as a buoy solution if the use case either permits a strong degree of control on the maximal distance between buoy and shore line (e.g. fjords, rivers or moored installations) or if the expected stranding area of the buoys is known as the radio then can work primarily as a recovery aid and the measurements can then be recovered from the onboard SD card. To maximize the chances of a successful radio transmission between buoy and base station, we developed a radio transmission protocol aimed at both ensuring that the buoy-base station connection is likely good enough for a data transfer to be successful while simultaneously ensuring that all received measurements are correctly paired with the buoy that took the measurements. The radio protocol is illustrated in Figure \ref{fig:protocol_diagram}, and is as follows:}
\begin{itemize}
    \item[\textbf{Step 1:}] The base station transmits its ID  and listening frequency $f_{b,r}$ at a predefined transmission frequency $f_{b,s} = \unit[868]{MHz}$.The listening frequency transmitted to the base station is unique for each base station, to avoid possible data duplication by having more than one base station receive data from the same buoy simultaneously. 
    \item[\textbf{Step 2:}] The base station switches to a listen mode at its own unique listening frequency frequency $f_{b,r}$. If a  buoy manages to catch the transmitted ID and has a sufficient amount of data for transmission, it will respond by transmitting a \emph{ready} signal consisting of the base station ID and the buoy's unique ID.
    \item[\textbf{Step 3:}] If the base station receives a \emph{ready} signal from a buoy which contains the base station's ID, it will  transmit  a \emph{go ahead} message at $f_{b,s}$ containing the base station ID, the buoy ID and the amount of time that the base station will listen for messages from the buoy. The buoy will read back the \emph{go ahead} message from the base station and start to transmit if the base station and buoy IDs both match the handshake transmitted in Step 2. 
    \item[\textbf{Step 4:}] The buoy transmits data to the base station until the buoy either has transmitted all of its available data or the allotted listen time (minus a certain grace period to avoid an overshoot) from the base station runs out. 
    \item[\textbf{Step 5:}] After the buoy has finished transmitting  data within the allotted listening time, the buoy  transmits an \emph{end} message before switching the radio to listening mode. This message is read by the base station, which then allows the base station to transmit new directives to the buoy. These directives include updates to the mutable parameters in Table \ref{tab:config_params}, or requests for packages that were lost in transit. As each measurement has a unique measurement ID and is stored on the onboard SD card, the base station only needs to transmit the ID of the requested measurement to the buoy for it to retransmit the measurement.
\end{itemize}

The protocol ensures that the buoy will not try to send any data (which is then, upon successful transmission, deleted from system memory) unless it has received a recent confirmation of having a base station within listening range. The transmitted data are encoded as a raw byte array before transmission to minimize the amount of data needed to transmit a single measurement. \revised{In the case of a message getting lost in transmission, the measurement data are still stored in the onboard SD card of the buoy.}

\section{Validation}

To validate the ORB framework, we performed a number of experiments with the ORB framework.
First, 
to determine the range limits of the distance between the base station and the buoy, we performed range experiments.
These  were performed in the ocean or across lakes, to ensure that our experiments closely resemble the in situ conditions that the buoys will encounter.
Second, to determine the buoy's ability to continue to operate in water, we performed reliability experiments.
These experiments demonstrated that the ORB buoy can operate for several months in ocean conditions. 

\subsection{Transmission Range Experiments}

\paragraph{Setup.}

We performed three experiments with the aim of establishing an estimate on the effective range of the LoRa communication between a base station and a buoy. Two experiments were performed outdoors near a lake with one or more buoys and one base station. The buoy(s) were mostly configured  with the standard configuration, but to make the experiments practical they were given  a reduced measurement period $T_m = 15$ s and a minimum transmission period $T_t = 15$ s. The measurements per transmission were $N_m = 4$. Hence, the buoy would try to send a message roughly every minute. In each experiment, the base station was placed at a known location while the buoy was carried around the lake.
During the first range experiment, at \emph{Sognsvann} lake, the buoy position at the time of transmission was validated using a Garmin Forerunner 255s watch. In combination with the timestamp of the message received from the base station, this ensured that we were able to compute the exact distance between the sender and receiver.
The second range experiment, at \emph{Nøklevann} lake, focused on transmission  distance. The experiment was only performed at two discrete distances due to the heavy foliage and distance between the road and the lake for most of the trip.
The third range experiment, in the \emph{Oslofjord},  was performed with a base station placed at a vantage point on land while the drifter was pulled by a boat in a straight line until we lost contact with the base station. In all experiments, the handshake was enabled to ensure that transmissions would occur only if the buoy was within range of the base station.

\paragraph{Results.}

\begin{figure}[t]
    \centering
   \input{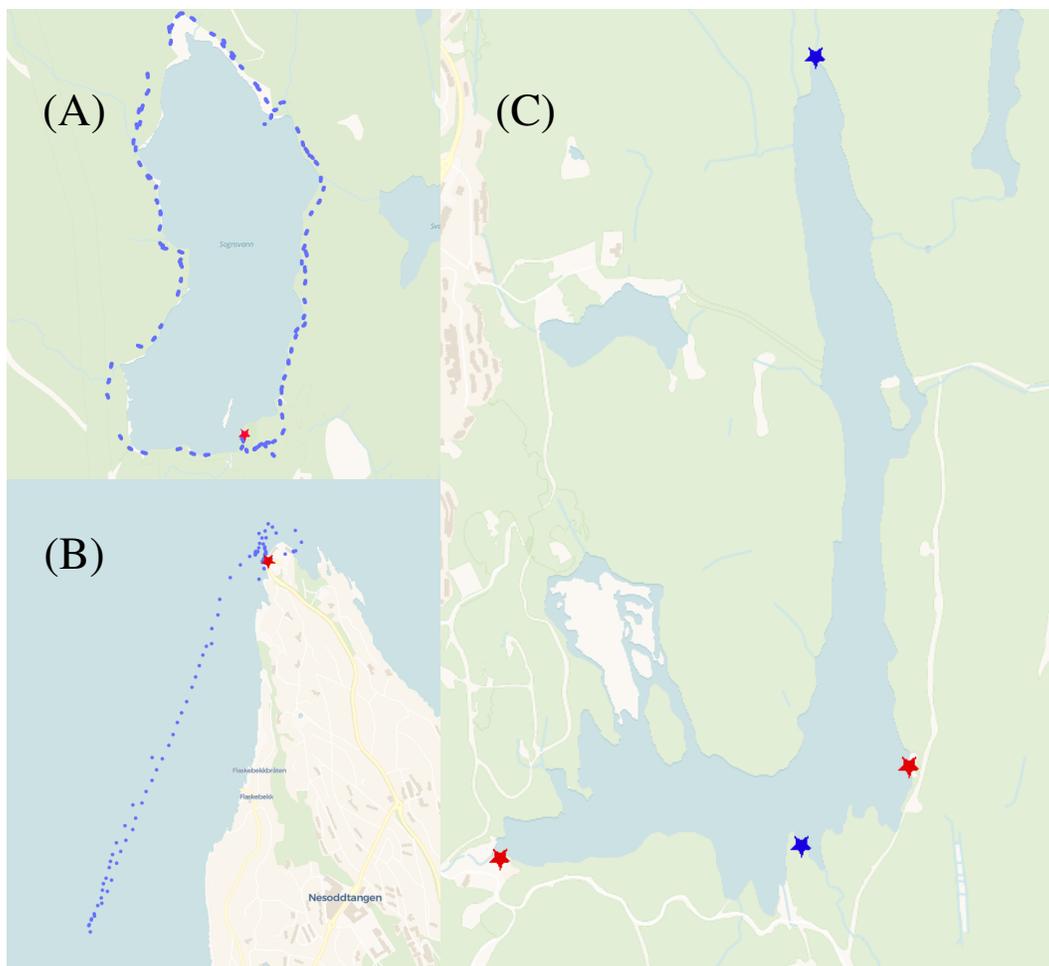}
   \caption{The performed transmission range experiments. Subfigure (A) shows the position of the buoy at the time it managed to establish a connection with the base station during the \emph{Sognsvann} range experiment. Each blue dot shows the measured position of the exercise watch at the timestamp of a confirmed transmission from the carried buoy. The position of the base station is marked by a red asterisk. The longest measured transmission distance was $\sim 1.2$ km. Subfigure (B) shows the location of the sender/receiver at the two sampling points at \emph{Nøklevann}. The distance between the the two red asterisks (located at the south east and south west of the lake) is 1.3 km, while the distance between blue asterisks (located at the southern and northern part of the lake) is 2.3 km. Subfigure (C) shows the trajectory taken by a buoy pulled by a boat in the \emph{Oslofjord}, the base station location is marked by a red asterisk.\label{fig:range_experiments}}
\end{figure}

 The connection was maintained throughout the \emph{Sognsvann}  experiment, with some exceptions in areas with dense foliage. The coordinate of each transmission is shown in Figure~\ref{fig:range_experiments}A. In the figure, the blue dots show the closest position in time of the exercise watch to a successful transmission from the buoy. The longest observed transmission range was $\sim 1.2$ km. In the \emph{Nøklevann} experiment, we were able to get a stable transmission between the buoy and the base station along both the short side  and the long side  of the lake. The positions in this experiment are shown with blue (long side experiment) and red (short side experiment) asterisks in Figure~\ref{fig:range_experiments}B. The distance between the sender and receiver was 1.3 km along the short side and 2.3 km along the long side of the lake, although the base station had to be placed in a manner that minimized the amount of foliage between the two. 
 In the \emph{Oslofjord} experiment,  we observed a maximum range of 1.8 km before communication was lost, as shown in Figure~\ref{fig:range_experiments}C. The results are summarized in Table~\ref{tab:ranges}.

\begin{table}[t]
    \centering
    \begin{tabular}{lcc}
      \toprule
      \textbf{Location} & \textbf{Max. range [km]} & \textbf{In water}  \\
        \midrule Sognsvann & 1.2 & No \\ 
         Nøklevann & 2.3 & No \\
         Oslofjord & 1.8 & Yes \\ 
         \bottomrule
    \end{tabular}
    \smallskip
    \caption{Observed ranges from the different range experiments. The ``In water'' column indicates if the buoy was floating in water or hand held during the experiment. }
    \label{tab:ranges}
\end{table}


\subsection{Buoy Longevity and Reliability Experiments}

\begin{figure}[t]
        \centering
        \includegraphics[width=0.8\linewidth]{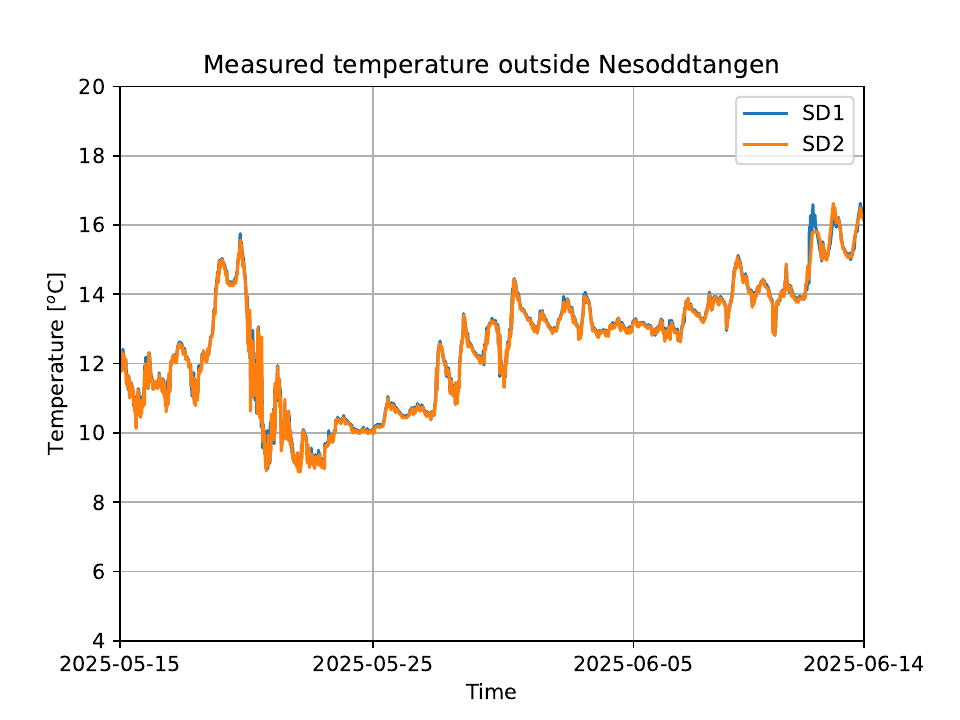}
        \caption{Temperature measurements outside of Nesoddtangen, recovered from two onboard SD cards. The buoys were deployed on May 15, 2025, and moored to an underwater installation, hence being subjected to oceanic conditions for the entirety of the measurement period.\label{fig:relibability_experiments}}
\end{figure}

\paragraph{Setup.}
We performed two experiments to validate the longevity and reliability of the ORB design. The first experiment focused on the buoy's reliability faced with long-term exposure to oceanic conditions, while the second experiment addressed the stability of the communication between each component. For the first experiment, we prepared two buoys with  default configuration parameters (shown in Tables \ref{tab:config_params} and \ref{tab:config_flags}). The two buoys were then moored to a communal underwater garden at Nesoddtangen.
A base station was placed on a vantage point nearby (approximate distance 30 meters). An image from the deployment of an earlier prototype in the same location is shown in Figure~\ref{fig:relibability_experiments}B. The buoys were moored by attaching a rope to the top rope of the underwater garden at two meters depth, and deployed on  Thursday May 8, 2025. 
The second experiment was performed starting on October 2, 2025. Here, one buoy was initialized with a low measurement period $T_m = \unit[20]{s}$,  the other parameters being the default values from Tables \ref{tab:config_params} and \ref{tab:config_flags}. This buoy was placed in a laboratory and a base station with the default parameter configuration  was placed beside the buoy. After two weeks, the buoy was powered off for three hours before being it was repowered, and two hours later the base station was powered off for three hours before being repowered. 

\paragraph{Results.}
The first reliability experiment, where two buoys were moored outside Nesoddtangen, demonstrated that the buoys functioned for a months time.  The recorded temperature by both buoys is shown in Figure \ref{fig:relibability_experiments}. The experiment did suffer from frequent loss of connection between the base station and the cloud due to an issue in the notecarrier in which it would enter sleep mode after a certain amount of time. This issue motivated the second test of the connection stability, which was performed after this issue had been solved. The second laboratory test found no signs of the base station entering an undesired sleep mode, with a stable connection for more than two weeks (data not shown).

\subsection{Testing the Interaction Between Multiple Buoys and Base Stations}
\paragraph{Setup.}
We performed a test with the aim of checking if several base stations and buoys would interfere, and if the buoys would exhibit some sort of base station preference. To this end, we deployed two base stations on opposite ends of \emph{Sognsvann} lake. Two buoys were activated, with a packet count threshold of 3, and measurement period $T_m = 15$ s. The two base station locations were placed  approximately 1.2 km from each other. The buoys were then walked from base station A to base station B alongside the eastern side of the lake, with one thermometer measuring the air temperature and the other being held in hand under a glove, ensuring a measurable difference in temperatures. Once the buoys were by base station B, the researcher at base station A picked up the base station and carried it back to base station B along the western side of the lake. 

\paragraph{Results.}
The messages from the buoys did not interfere or get mixed up by the base station, as the \textit{warm} and \textit{cold} buoys were at no point mixed up in the signals (data not shown). Furthermore, there is little discernible difference in preference between the two base stations, showing that there is no implicit discrimination of base stations from the buoys end. Hence, the handshake acts as the check to see if the connection between buoy and base station is sufficient for communication, as was expected.

\subsection{A Live Deployment in the Drammensfjord}
\begin{figure}[t]
    \centering
    \input{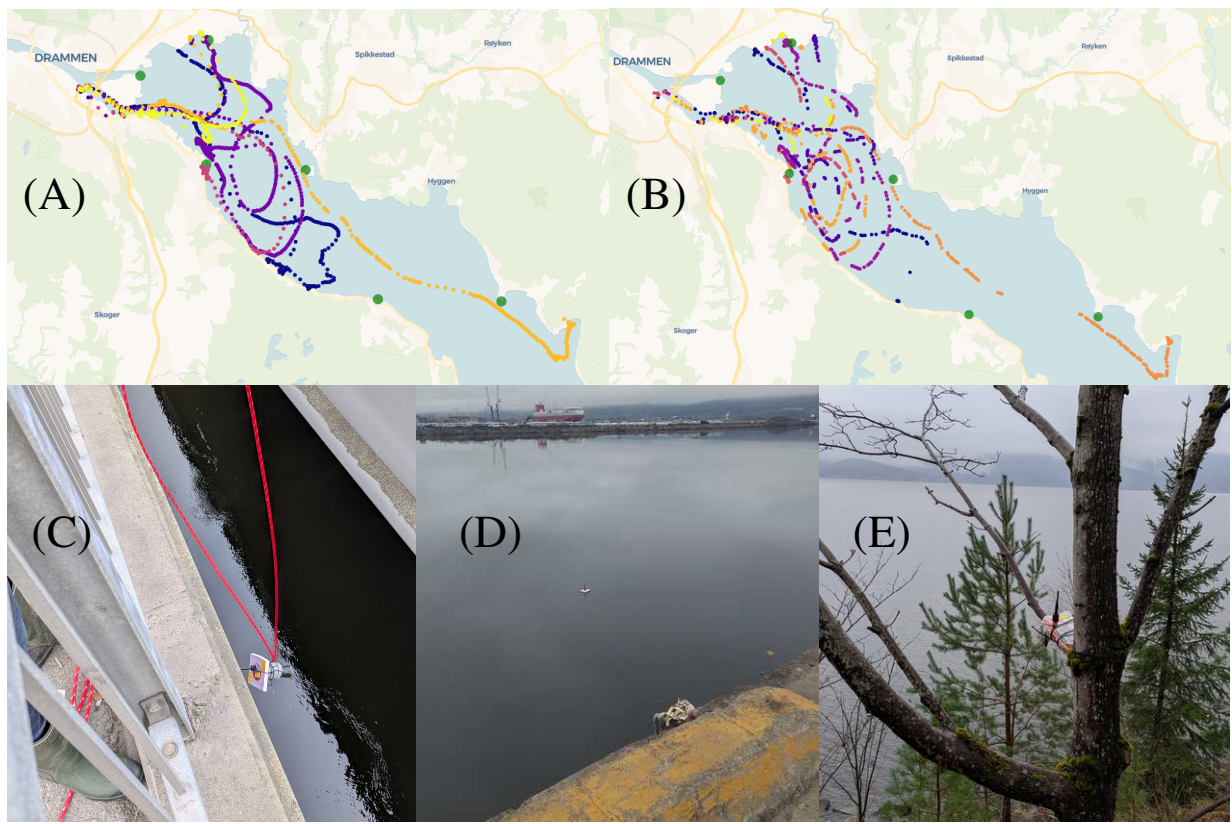}
    \caption{The observed trajectories from the \emph{Drammensfjord} deployment, alongside some illustrations of the deployment. Subfigure (A) show the recovered trajectories from the onboard SD card, while Subfigure (B) show the buoy coordinates that was received by a base station. The trajectories are colour coded by buoy ID. Subfigures (C) to (E) show the deployment of a buoy into the Drammen river (C), a buoy drifting into the fjord (D) and a base station at a vantage point (E).}
    \label{fig:Drammen}
  \end{figure}

\paragraph{Setup.}
We assembled 16 ORB buoys and 8 base stations that were used for  a first field experiment in the \emph{Drammensfjord}\revised{. The aim of the experiment was twofold, we wanted to both test the ORB solution in field conditions while simultaneously measuring and collecting data on the  fjord circulation as the ecological conditions in the entirety of the greater Oslo fjord is critical} \citep{OsloFjordRapport}. The base stations were deployed at the locations marked with a green marker in Figure \ref{fig:Drammen}A and B. The buoys were deployed from a bridge at the southern bifurcation of the Drammen river close to the estuarine outlet. The buoys were coded with the standard parameter configuration in Table \ref{tab:config_params}, except for a reduced measurement period $T_m = \unit[5]{min}$. Figure \ref{fig:Drammen}C shows the deployment procedure, Figure \ref{fig:Drammen}D a buoy drifting into the fjord and Figure \ref{fig:Drammen}E a base station deployed at a strategic vantage point close to the fjord. 

\paragraph{Results.}
Of the 16 deployed buoys, we were able to recover the remote particle trajectories of 12  buoys (either in full or in part). Furthermore, we were able to recover 8 of the deployed buoys, with the remaining 8 either having lost signal before hitting shore, or hitting shore at locations that were too difficult to reach for a safe recovery to be possible. From the recovered buoys, we were able to obtain the raw data from 7 of the 8 onboard SD cards. The obtained SD card trajectories are shown in Figure \ref{fig:Drammen}A, while the transmitted trajectories are shown in Figure \ref{fig:Drammen}B. The maximal measured transmission distance during this deployment was 5.3 km.


\begin{figure}[t]
    \centering
    \includegraphics[width=0.8\linewidth]{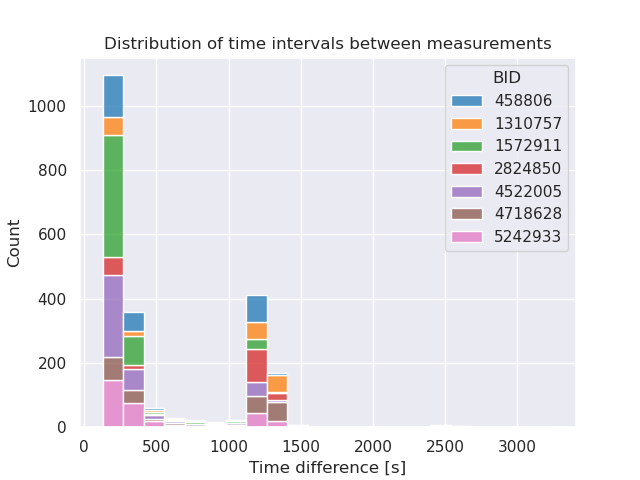}
    \caption{The time interval distribution of GPS measurements. Each colour represent one recovered buoy, with the shown data being retrieved from the onboard SD card. Four time intervals were above one hour, and have been cut from the figure for legibility. }
    \label{fig:GPS_histogram}
\end{figure}

The majority of the measurements ($\sim$\,70 \%) were taken in the desired $T_m= \unit[5]{min}$ interval, but a significant number ($\sim$\,30 \%) were taken with a large time interval in between measurements. The distribution of these measurements is shown in Figure \ref{fig:GPS_histogram}. Different buoys exhibited clear differences in reliability, most notably Buoy 1572911 measured predominantly on time, while Buoy 2824859 only sporadically measured on time, with the majority of GPS fixes taking close to about 20 minutes to get, suggesting that care should be made during assembly, as human error might adversely affect buoy performance. 

\section{Discussion and Conclusion}

The ORB framework is a complete system package to measure oceanographic flows in coastal and fjord environments. A base station can handle multiple buoys during a given deployment, and the battery lifetime has been shown to exceed three months in the field. Furthermore, the comparatively low price of 100 USD/buoy makes it possible to acquire many buoys for a field experiment within a constrained budget. Hence, the ORB would be suited for capturing spatial variance in flow patterns, as well as getting an estimate for ensemble statistics of the measured flow. The greatest limitation of the ORB is its transmission range, with an observed upper bound of 5 km, which might be lower in the presence of waves. This limitation makes the ORB less suitable for measurements in the open ocean far from the coast, as it would be challenging to obtain sufficient coverage from the base stations. A moored installation further offshore should be possible to construct, following the example of \citet{raphael2025low}.

The ORB buoy as a device can perform similar measurements as previously developed buoys, such as \citet{da2022generation, majumder2024buoys}, \citet{raphael2025low} and \citet{smartOcean}. However, the ORB buoy  stands out in its open-source aspect. The source code, hardware, and data platform are publicly available and modifiable under a permissive license. Open-source instrumentation has several distinct advantages. Similarly to, e.g. the OpenMetBuoy \citep{rabault2022openmetbuoy}, the SFY \citep{hope2025sfy} and microSWIFT \citep{thomson2024development}, the ORB buoy is fully usable out of the box at the cost price of assembling the buoy. Furthermore, since both the code and hardware are openly available, open buoys such as ORB also serve as a solid starting point for users whose needs are almost or somewhat covered by the existing designs. Open-source instruments are also able to strengthen their own source code and hardware through community feedback, since all relevant information is publicly available, strengthening confidence in the solution in the long term.

The ORB buoy has been shown to be stable over long time periods and to be energy efficient. With more than three months (in the worst case scenario) of continuous operation on a single 13 Ah lithium battery, we can reasonably assume that the ORB buoys have sufficient battery lifetime to last for most drifter-in-fjord deployments due to the requirement for the buoys to be close to a coast with mounted base stations. For moored and long-time deployments, additional battery capacity can be obtained simply by adding more batteries to the buoy. Several power consumption optimizations have been applied to extend battery lifetime, and we are continually working to improve this aspect further where possible. In combination with the moderate price tag of $\sim$ 100 USD, this means the ORB buoy is a viable low-cost alternative to commercial and closed-source buoys if the targeted use case lies within the range constraints of LoRa technology. 

\revised{The field of open-source instrumentation is crowded, and a new design should either seek to complement the existing and maintained technologies, or prove to be an undeniable improvement to an existing solution. The OpenRadioBuoy aims to be the former, seeking to supplement buoy systems such as the SFY \citep{hope2025sfy}, the OMB \citep{rabault2022openmetbuoy} and the microSWIFT \citep{thomson2024development}. Most existing buoy systems rely on either GSM or Iridium for their communication protocol. These options, while permitting greater range and lower weather sensitivity \citep{elijah2021effect}, come with the drawback of higher power use, the SFY take about an average of \unit[130]{mA} during transmission and the OMB about \unit[250]{mA} \citep{rabault2022openmetbuoy, hope2025sfy}. Some buoy solutions come with  data transfer costs as well, for example the OMB came with Iridium transmission costs of about \unitfrac[100]{USD}{month} at the time of publication \citep{rabault2022openmetbuoy}. These costs to power and research funds are often necessary, given the individual challenges of the phenomena a researcher wish to study. The wide span in different research scenarios and their respective challenges have resulted in the academic community not converging on a single ''one buoy measures all'' solution, instead developing a wide variety of buoys meant to fit their respective niches  \citep{cavaleri2025more}, and we have given a non-exhaustive list of open instruments in Table \ref{tab:buoy_alternatives}.}

\revised{There are several situations where the ORB is not a suitable buoy solution, and where the aforementioned buoys are likely to be a preferable option. If, however the geometry of a study limits the maximum distance between buoy and land, as is the case with fjords, rivers or if the possible expected stranding area of where the buoys will hit land can be sufficiently covered, then we believe the ORB to be a suitable low-cost instrument for the study. The ORB is both inexpensive, can be assembled with off-the-shelf products and is easily configurable without major changes to the source code. Finally, the ORB can be relatively easily modified to be installed as a more permanent moored sensor platform for near-coastal structures, where the low energy use of the radio protocol can ensure a long battery life before having to change the batteries thereby offering a cheap and low-maintenance sensor platform.}

The ORB buoy, as described in this paper, is an open instrument that is still under development, and extensions to different sensors and use cases are developed. For example, work has already been done on a turbidity sensor library, with applications to permanent moored sensors in mind. Furthermore, all data that we gather using the ORB buoy will also be made publicly available. This will allow for easier validation of coastal ocean models in general and of fjord environments in particular. The comparatively low price allows for a greater statistical representation of the surface drift, making it easier to validate a numerical model against observed patterns. Furthermore, with the increasing number of digital twin infrastructure initiatives for ocean monitoring (e.g. \citet{chen2023toward, tzachor2023digital}) and work with trace DNA (e.g. \citet{shelton2022environmental}), this information is also valuable outside of oceanography.

\section*{Acknowledgements}

This work was supported by the Sustainability initiative of the University of Oslo and the Digital Arctic Twins network (DART, funded by the University of the Arctic). 
The authors thank Lars Dalen and Marinreparatørene for inviting us to use their communal underwater garden at Nesoddtangen for validation experiments, as well as for helping us to deploy and recover the buoys. We also thank Issam Rais at the Arctic University of Tromsø for his help and code review during a hackathon at Marinreparatørene, and Olav Gundersen at the Hydrodynamical Laboratory at the Department of Mathemathics at the University of Oslo for his assistance in the workshop with buoy construction. A special thanks goes to Ivar Vannebo and Morten Hansen at Drammen Havn and Hans Martin Hansen for letting us install base stations at their premises, and assisting us with the \emph{Drammensfjord} experiment. Finally, we thank Daniel Sæther and Tor-Erik Kristiansen for assisting us in recovering the buoys once they had drifted ashore.

\bibliographystyle{plainnat}
\bibliography{bibliography}

\section*{Appendix A: Open-Source Release}
All firmware for the buoy, base station and digital platform will be made available at \url{https://github.com/larswd/OpenRadioBuoy} after publication in the peer-reviewed literature at the latest. This repository also contain the printed circuit board (PCB) designs for the buoy and base station, alongside a maintained component list with links to a distributor. 
\end{document}